\documentclass[final]{svjour2}
\usepackage{graphicx}
\usepackage{rotating}
\usepackage{amssymb}
\usepackage{mathptmx}
\usepackage[numbers]{natbib}
\usepackage[]{amsmath}
\usepackage{mathrsfs}
\usepackage{epsf}
\usepackage{psfrag}
\usepackage{dcolumn}
\usepackage{bm}
\usepackage{setspace}
\usepackage[geometry]{ifsym}
\usepackage[dvips]{color}
\usepackage{subfigure}
\makeatletter
\journalname{Journal of Low Temperature Physics}

\bibpunct{}{}{,}{s}{}{,}

\begin{document}

\newcommand{\hdblarrow}{H\makebox[0.9ex][l]{$\downdownarrows$}-}
\title{A Fabrication Route for Arrays of Ultra-Low-Noise MoAu Transition Edge Sensors on thin Silicon Nitride for Space Applications.}
 
\author{D. M. Glowacka $^1$ \and M. Crane$^1$ \and D. J. Goldie$^1$ \and S. Withington$^1$}

\institute{1:Detector and Optical Physics Group, Cavendish Laboratory,\\ University of Cambridge,
J. J. Thomson Avenue, Cambridge, CB3 0HE, UK\\
Tel.:+1223 337366\\
\email{d.m.glowacka@mrao.cam.ac.uk}
}

\date{18.07.2011}

\maketitle
\begin{abstract}

We describe a process route to fabricate arrays of Ultra-Low-Noise MoAu Transition Edge Sensors (TESs). The low thermal conductance required for space applications is achieved using 200 nm-thick 
Silicon Nitride (${\rm SiN_x}$) patterned to form long-thin legs with widths of $2.1\,\,{\rm \mu m}$.  Using bilayers formed
on ${\rm SiN_x}$ islands from films with 40 nm-thick Mo and Au thicknesses in the range 30 to 280 nm deposited by dc-sputtering in ultra-high vacuum we can obtain tunable transition temperatures in the range 700 to 70~mK. The sensors use large-area absorbers fabricated from high resistivity, thin-film $\beta$-phase Ta to provide impedance-matching to incident radiation. The absorbers are patterned to reduce the heat capacity associated with the nitride support structure and include Au thermalising features to assist the heat flow into the TES. Arrays of 400 detectors at the pixel spacing required for the long-wavelength band of the far-infrared instrument SAFARI are now being fabricated.  Device yields approaching $99\%$  are achieved. 

\end{abstract}

\section{Introduction}
\label{sec:Intro}

The far-infrared (FIR) band from $300\,\,{\rm \mu m}$ to $20\,\,{\rm \mu m}$ is of crucial importance 
for astrophysics, particularly with regard to studying the formation and evolution of galaxies, stars and planetary systems.
 It encompasses the wavelength range of the planned space missions 
SPICA\cite{Swinyard_1},  FIRI\cite{FIRI} SAFIR, SPECS, and SPRIT; 
and it covers the wavelength range of the balloon interferometers BETTII\cite{BETTII}
 and FITE.\cite{FITE}
 The new generation of space missions covering this range will require large-format arrays of extremely low noise detectors with Noise Equivalent Powers (NEPs) of order $ 10^{-19}\,\,{\rm W/\sqrt{Hz}}$ or better.
This requirement has driven the development of sensitive superconducting detectors, such as Transition Edge Sensors (TESs). 
While it is technically feasible to manufacture single TESs having this sensitivity, it is challenging to create an ultra-low-noise TES technology that can be engineered into complete imaging arrays, 
with the required optical packing and uniformity of performance. 
%
%
\begin{figure}[ht]
\centering
\psfrag{f in Hz}[] [] {Frequency (Hz)}
\psfrag{yaxis}[] []{$NEP \left( \times 10^{19} {\rm W/\sqrt{Hz}}\right) $ }
\subfigure[$\qquad\qquad\qquad\qquad\qquad\qquad$  ]{
\includegraphics[height=5 cm, angle=0]{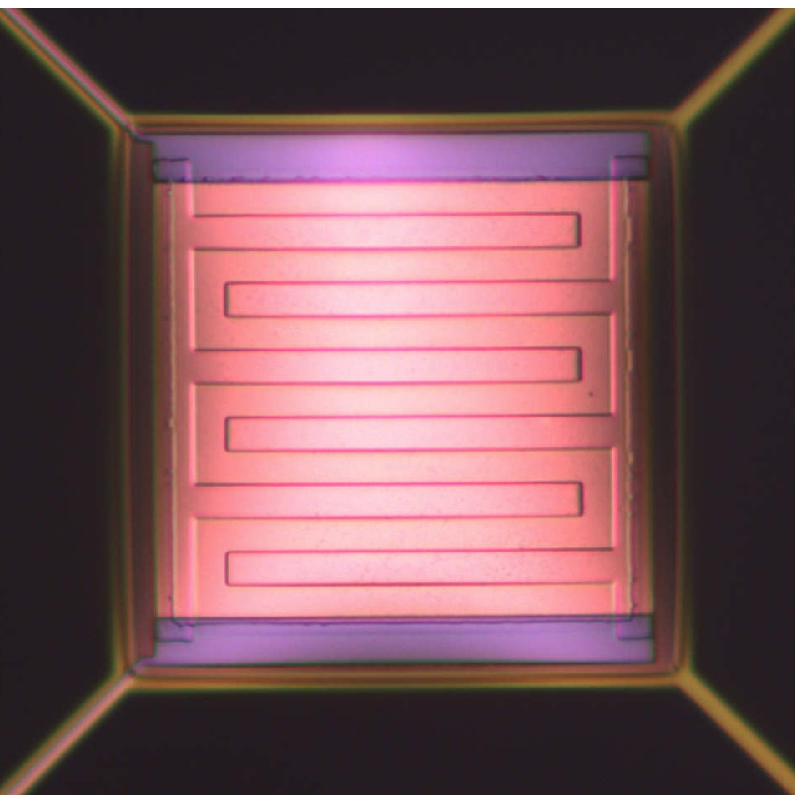}
\label{fig:Mo_Cu_TES}
}
\subfigure[$\qquad\qquad\qquad\qquad\qquad\qquad$ ]{
\includegraphics[height=5 cm, angle= 0]{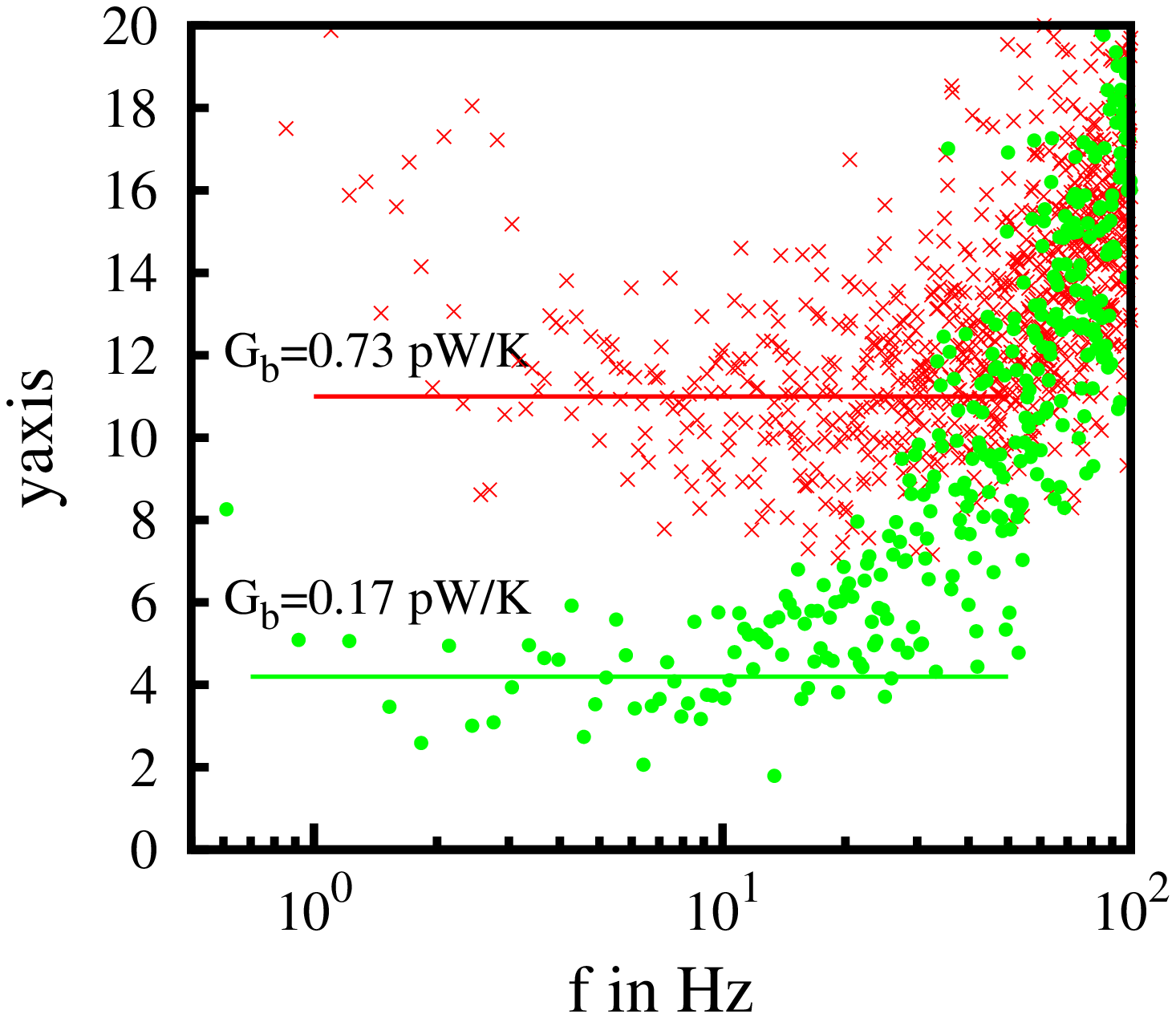}
\label{fig:MoCu_darkNEP}
}
\label{fig:Earlier_Device}
\caption[Optional caption for list of figures]{
(Color online) \subref{fig:Mo_Cu_TES}
Optical image of an earlier MoCu TES. 
\subref{fig:MoCu_darkNEP} 
Dark NEP for an earlier MoCu TES reported in Ref.~[7].
}
\end{figure}
The NEPs required for these sensitive detectors can be achieved by thermally isolating
the TES using four long ($\ge 500\,\,{\rm \mu m}$) and narrow ($\le 4\,\,{\rm \mu m}$) support legs 
fabricated using  thin ${\rm SiN_x}$. 
We have previously made and characterized
small arrays of MoCu TESs with $T_c\sim100\,\,{\rm mK}$ both with and without IR absorbers 
 as part of a development effort
to move from microstrip-coupled polarization-sensitive TESs with
$NEP\sim 2\times10^{-17}\,\,{\rm W/\sqrt{Hz}}$ tailored for ground-based
astronomy\cite{Damian_ISSTT,Dorota2008}
to the ultra-low-noise performance required for space applications.\cite{Goldie_trp1,Dorota2010}
 Figure~\ref{fig:Mo_Cu_TES}
shows an image of one of these MoCu
 devices fabricated without an additional  absorber structure. 
The curvature in the central ${\rm SiN_x}$ island is of order $1\,\,{\rm \mu m}$ out of the figure and arises
from residual stress in the ${\rm SiO_2}$ layer needed to passivate the Cu. The  ${\rm SiO_2}$
 also contributes  heat capacity and hence possibly additional noise features. 
MoAu TESs do not require passivation and our initial devices reported 
elsewhere in these proceedings\cite{Goldie_MoAu1}  have demonstrated 
excellent long-term stability. 
Figure~\ref{fig:MoCu_darkNEP} shows the measured dark NEP 
for two of these TESs. The low frequency $NEP\sim 4\times10^{-19}\,\,{\rm W/\sqrt{Hz}}$  
for a conductance to the bath $G_b=0.17\,\,{\rm pW/K}$ is within a factor of order $\sqrt{2}$ of the phonon limit
calculated with noise modifier $\gamma_\phi=0.5$.\cite{Mather} 

In this paper we describe the fabrication route for ultra-low-noise MoAu TES detectors operating
close to $100\,\,{\rm mK}$ with a brief report on the status of the latest measurements.
\section{Experimental details}
\label{sec:Experimental Details}
\begin{figure}[ht]
\centering
\subfigure[ $\qquad\qquad\qquad\qquad\qquad\qquad\qquad$]{
   \includegraphics[height=3.5 cm, angle=0]{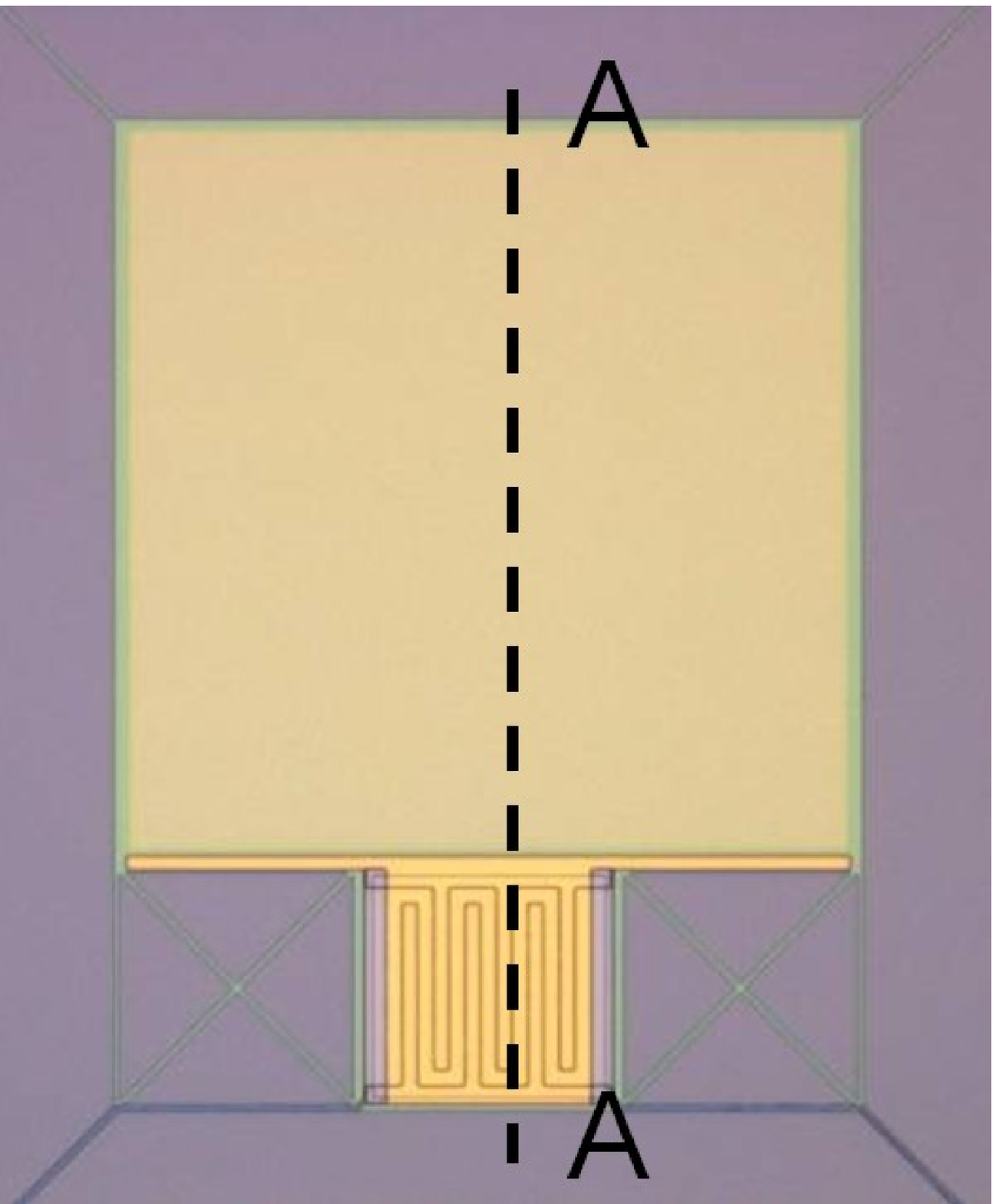}
\label{fig:fullabs}
}
\subfigure[ $\qquad\qquad\qquad\qquad\qquad\qquad\qquad$]{
   \includegraphics[height=3.5 cm, angle=0]{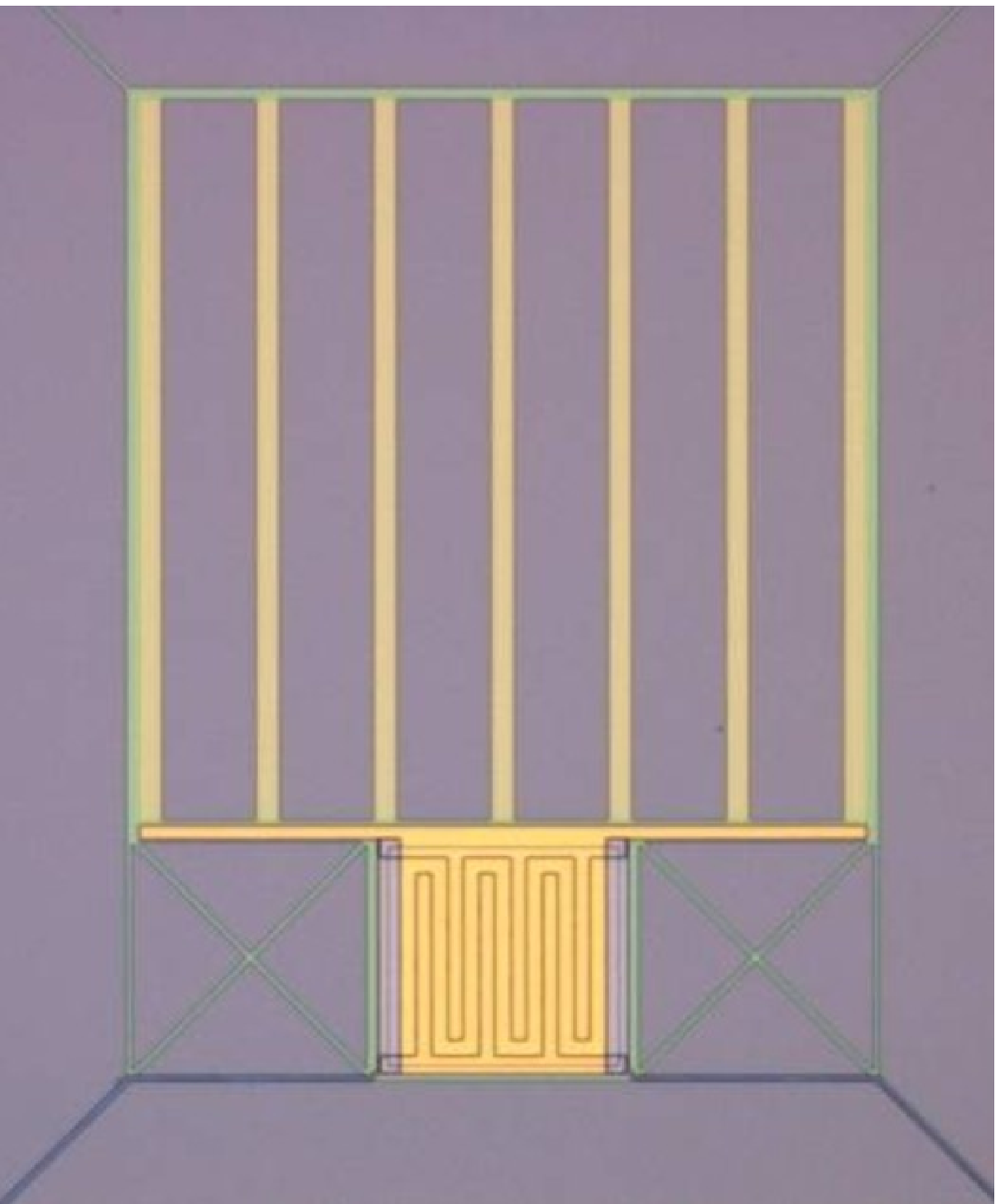}
\label{fig:striped}
}
\subfigure[ $\qquad\qquad\qquad\qquad\qquad\qquad\qquad$]{
   \includegraphics[height=3.5 cm, angle=0]{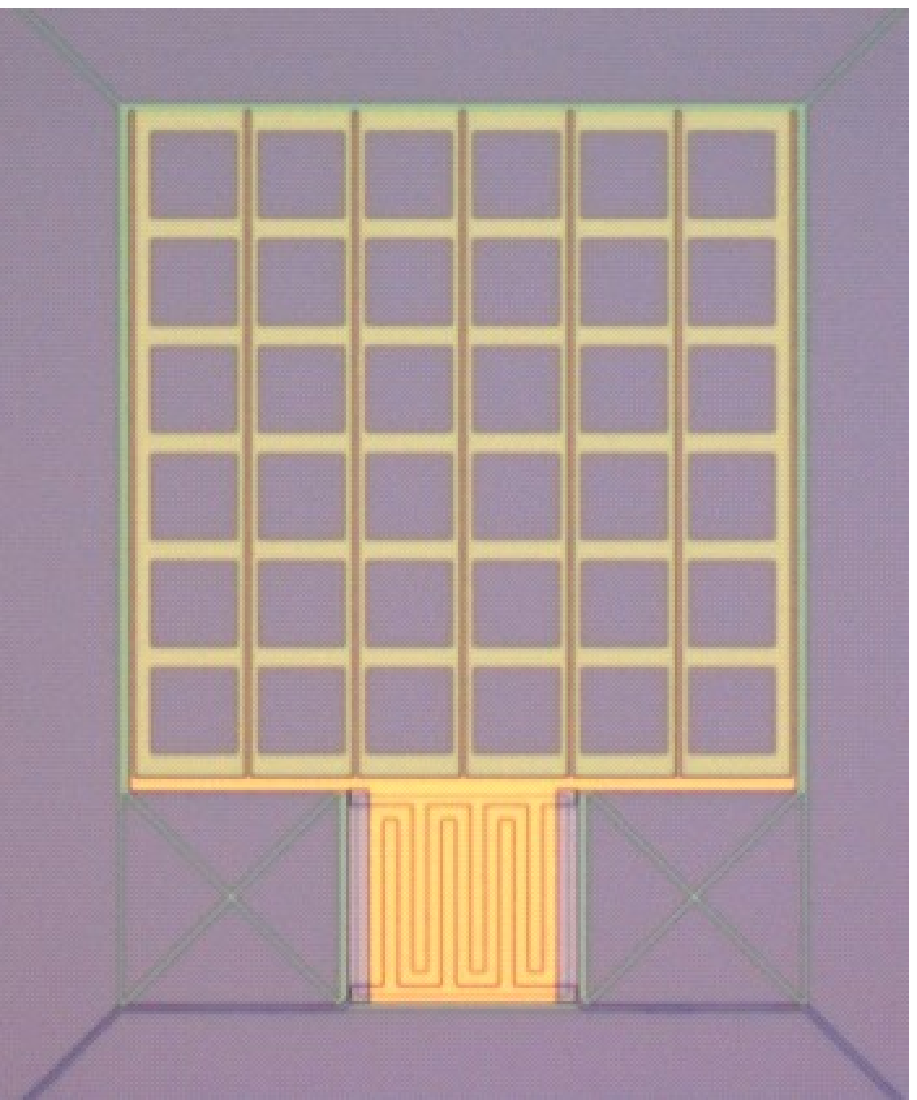}
\label{fig:meshed}
}
\label{fig:TES_pictures}
\caption[Optional caption for list of figures]{(Color online) Optical images of: \subref{fig:fullabs}  
A single $100\times 100\, {\rm \mu m^2}$ Mo/Au TES with longitudinal and partial lateral Au bars across the bilayer and Ta
absorber. The 200~nm thick ${\rm SiN}_x$ island under the TES has an area
$110\times 110\, {\rm \mu m^2}$. The Ta absorber is $320\times 320\, {\rm \mu m^2}$.
The supporting legs are $2\,{\rm \mu m}$ wide and $540\,{\rm \mu m}$ long.
\subref{fig:striped} 
A striped absorber with seven $25 {\rm \mu m}$-wide Ta bars and \subref{fig:meshed} a meshed Ta absorber patterned into a 
$25 {\rm \mu m}$-wide grid to investigate the effect on the heat capacity.
}
\end{figure}

The array described here has 388 single devices each consisting of a MoAu bilayer TES with Au banks 
and fingers partially across the TES,\cite{Ullom2004}
 a Ta FIR absorber, Au thermalizing structures
and Nb connections all formed on a ${\rm SiN_x}$ island.
 The island is thermally isolated from the Si 
wafer by four 
${\rm SiN_x}$ legs of width $2.1\,{\rm \mu m}$ and 
length $540\,{\rm \mu m}$, which determine the thermal
conductance and hence the responsivity, saturation power and readout noise spectrum of the device.
Photographs of the device types are shown in Fig.~\ref{fig:TES_pictures}.

   \begin{figure}[ht]
   \begin{center}
   \begin{tabular}{c}
   \includegraphics[width=9.5cm, angle=0]{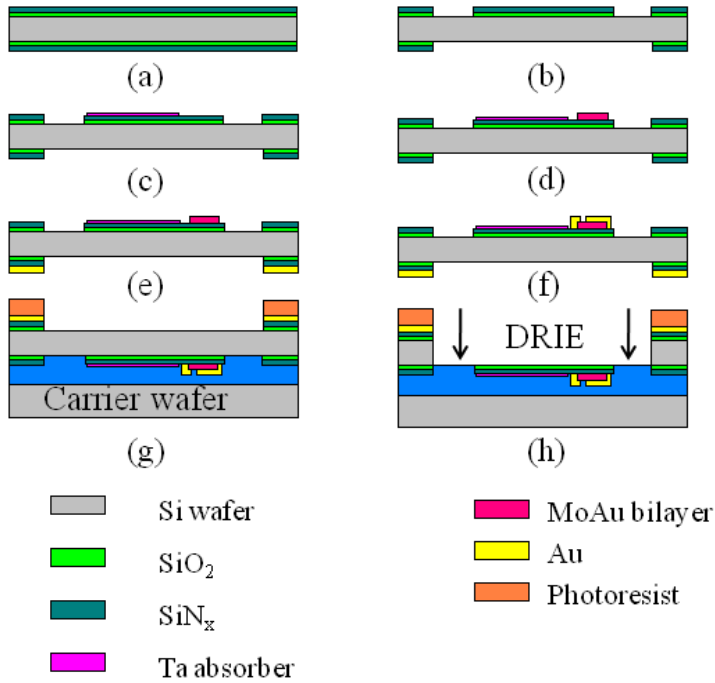}
   \end{tabular}
   \end{center}
   \caption[Fig2]
   { \label{fig:flow_diagram}
(Color online) Schematic diagram showing the process flow for a single array. The cross section is along the line 
marked A-A in Fig.~\ref{fig:fullabs}. }
   \end{figure}
%
The sensors are manufactured on 50~mm diameter, $225\,\,{\rm \mu m}$ thick 
$\langle 100\rangle$, 
double-side polished Si wafers. These wafers have, on both sides, 
a 50~nm film of thermal ${\rm SiO_2}$ and a 200~nm film spell of 
slightly off-stochiometric  ${\rm SiN_x}$, 
formed by low-pressure chemical vapour deposition. 
Ultimately, the ${\rm SiN_x}$ becomes a suspended membrane, 
produced by back-side etching of the supporting silicon substrate 
under the TES and absorber, to give the required thermal isolation. 

The first two steps of device fabrication are 
Reactive Ion Etching (RIE) of the ${\rm SiN_x}/{\rm SiO_2}$ using ${\rm CHF_3 }$
gas to outline the array and define the membrane and legs 
(done in two separate steps, one on the front of the wafer and one on 
the back): Fig.~\ref{fig:flow_diagram}(a) and (b). 
Then the Ta absorber layer is deposited by DC 
magnetron sputtering in a system with a base pressure of $10^{-10}\,\,{\rm Torr}$ as in
 Fig.~\ref{fig:flow_diagram}(c). 
The next step is deposition of superconducting bilayer by 
DC magnetron sputtering also under ultra-high vacuum conditions.
 The bilayer described here consists 
of a 40~nm Mo layer followed by a 195~nm Au layer, deposited 
in quick succession to avoid interface contamination and thereby 
achieving a reproducible proximity effect. The bilayer is patterned by 
wet etching: first the Au in a $50\%$
 solution of commercially 
prepared gold etch and water, and then the Mo in a commercially 
prepared aluminum etch: Fig.~\ref{fig:flow_diagram}(d).  
The next 
processing step is the deposition of a 200~nm-thick Au thermalizing layer 
on the back of the wafer: Fig.~\ref{fig:flow_diagram}(e).
A second 200~nm Au deposition 
forms thermalizing bars on the Ta absorber, and Au banks and lateral 
bars on the bilayer to improve thermalization of the TES: Fig.~\ref{fig:flow_diagram}(f).
A AuCu deposition, not shown, forms resistive heaters on the Si frame
and Johnson noise thermometers on selected ${\rm SiN_x}$ islands in place of the TES. 
The TES electrical connections and contact pads are formed from 
250~nm of Nb (not shown on the process flow). 
At each stage, the Au and Nb layers are removed from unwanted areas by lift-off. 
The wafer is then ready for bonding face down to a carrier wafer: Fig.~\ref{fig:flow_diagram}(g). 
The last step is the fabrication of the membrane, which requires removal of 
the supporting Si from the window using Deep Reactive Ion Etching, 
thus leaving the TES and absorber membrane suspended on the nitride legs: Fig.~\ref{fig:flow_diagram}(h).
Finally the wafer is demounted, cleaned in an 
${\rm O_2}$  plasma and is ready for testing.

The pixel spacing for the 
prototype array was matched to the SAFARI L-band
requirement  of 1.6~mm.
The array includes full Ta absorbers as in Fig.~\ref{fig:fullabs}, striped (Fig.~\ref{fig:striped})
and 
meshed absorbers (Fig.~\ref{fig:meshed}). 
On selected calibration pixels the Ta was omitted completely and on others 
both the absorber and its 
support nitride was removed. Johnson noise thermometers were also
 included to
assist with estimates of stray-light. The target $T_c$ 
was in the range $110-120\,\,{\rm mK}$.
The array was cooled in a closed-cycle, sorption-pumped dilution refrigerator 
mounted on a pulse-tube cooler giving a base temperature of 68~mK.\cite{Teleberg2008} 
The array  was enclosed in 
a Au-plated Cu box the inside of which was coated with light-absorbing
SiC granules and carbon black mixed in Stycast 2850 to minimize scattered light. 
A photograph of the experimental enclosure is shown in 
Fig.~\ref{fig:Mounted_device}. 
The sample space
was surrounded by multiple layers of Nb foil and Metglas to provide
magnetic shielding. We used  NIST SQUIDs  with
analogue electronics readout that keeps 
the multiplexer in a fixed state but
none-the-less permits readout of multiple channels.
For these tests the SQUIDs and bias resistors have a separate light-tight enclosure
with light-tight electrical feedthrough to the array enclosure. 
\section{Results and Discussion}
\label{sec:Results}
\begin{figure}[ht]
\centering
\subfigure[$\qquad\qquad\qquad\qquad\qquad\qquad$  ]{
   \includegraphics[height=4 cm, angle=0]{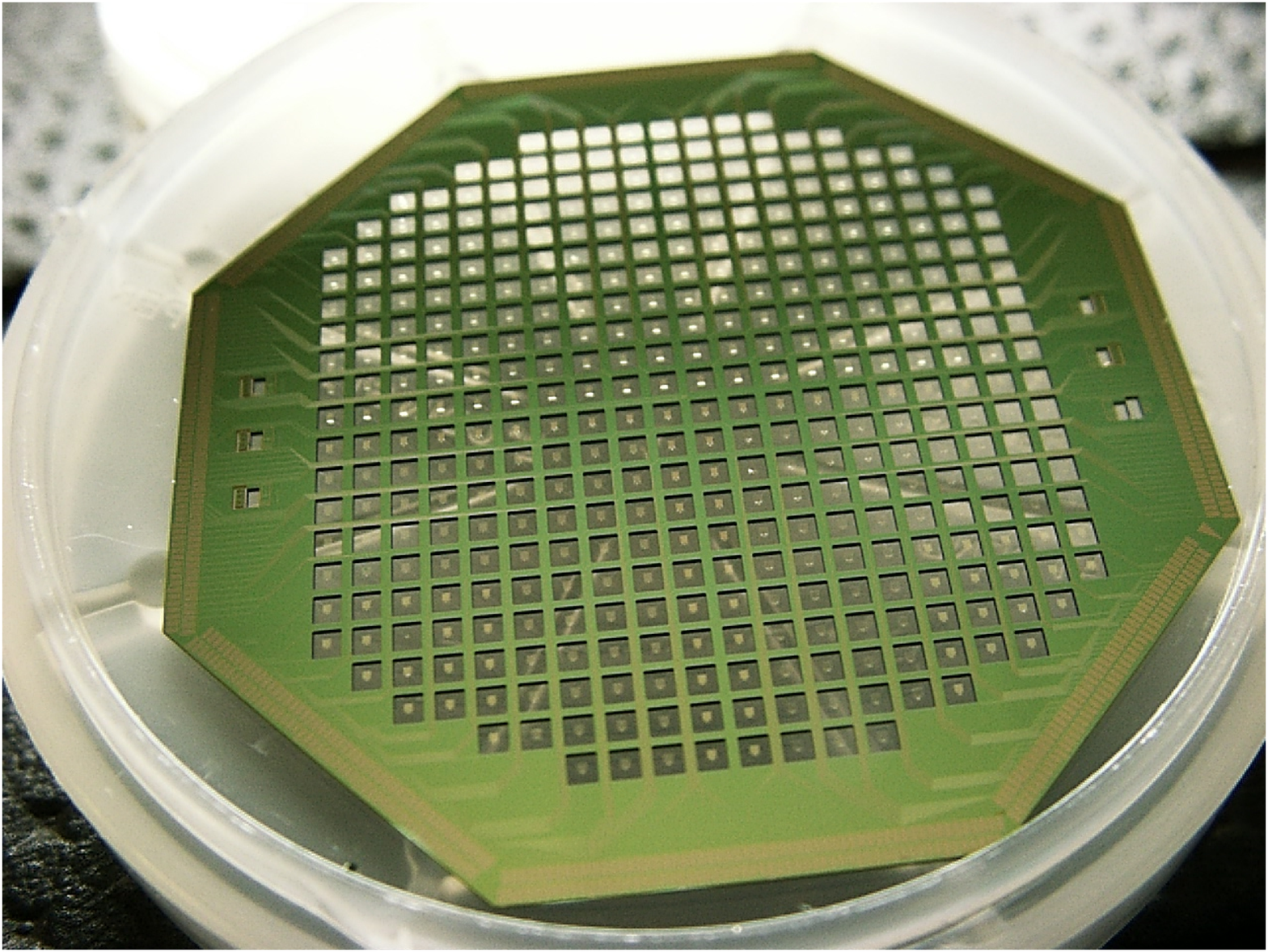}
\label{fig:array_chip}
}
\subfigure[$\qquad\qquad\qquad\qquad\qquad\qquad$ ]{
\includegraphics[height=4 cm, angle=0]{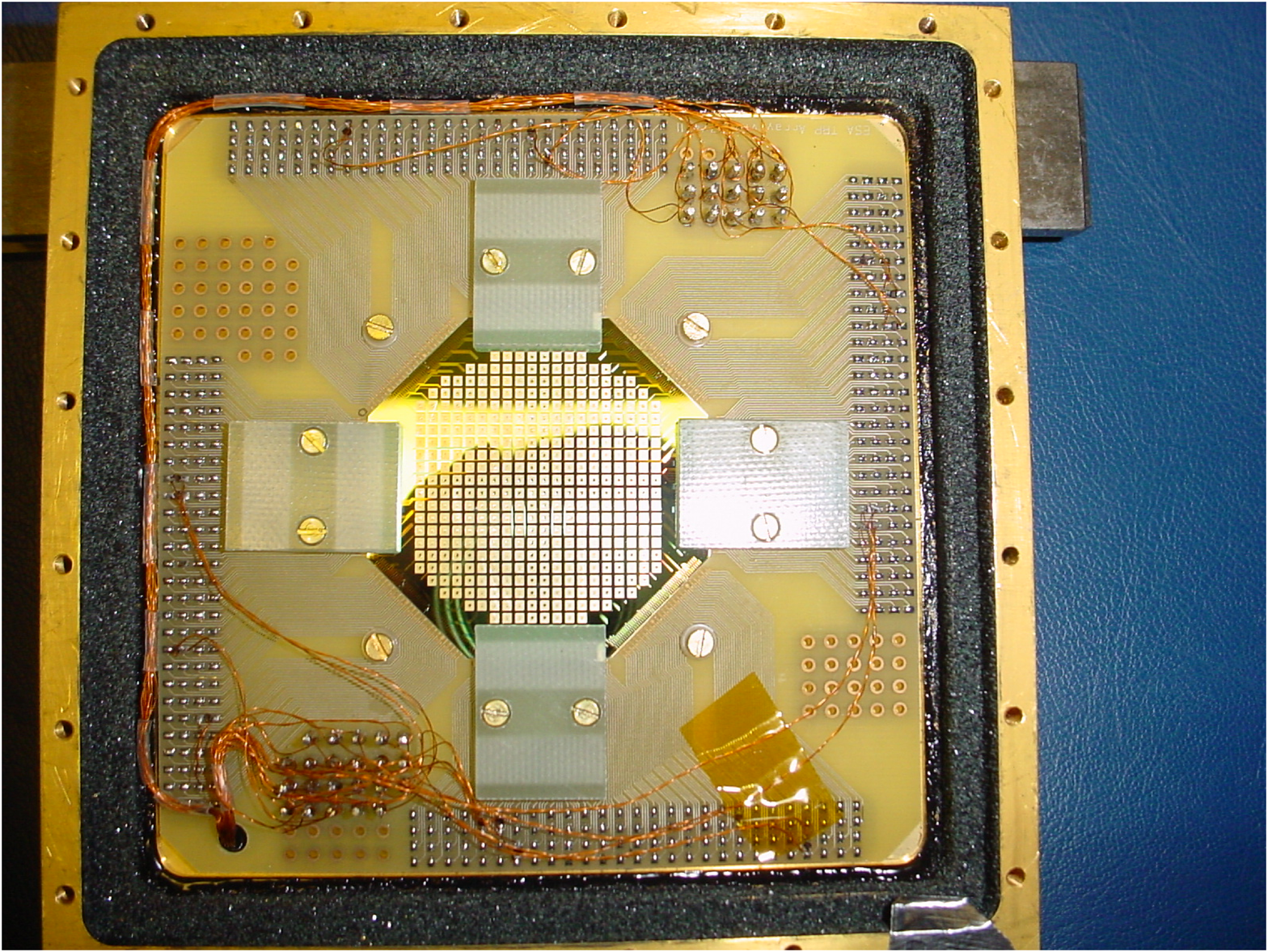}
\label{fig:Mounted_device}
}
\label{fig:Experimental_setup}
\caption[Optional caption for list of figures]
{
(Color online) \subref{fig:array_chip}
Completed array chip. 
\subref{fig:Mounted_device} 
Complete array mounted in a light-tight enclosure for low temperature testing.
}
\end{figure}
The first 388-element arrays have now been completed and 
low temperature measurements are in progress. 
Thermal cycle tests already indicate that the  mounting scheme shown in Fig.~\ref{fig:Mounted_device}
is robust and no devices have been damaged on repeated thermal cycles
despite the removal of a large fraction of the support Si (of order 36\%). 
Overall device yield
was 99\%. All TESs are clearly visible in Fig.~\ref{fig:array_chip} including in the lower left
portion a very small test pixel where the absorbing Ta and support ${\rm SiN_x}$ was removed  and the striped absorbers in the lower right sector. 
On a subset of TESs normal state resistances were measured to be $R_n=31\pm 1 \,\,{\rm m\Omega}$ and
$T_c=113\pm 5 \,\,{\rm mK}$. Conductance to the heat bath was $0.19\pm 0.2 \,\,{\rm pW/K}$ giving
a predicted phonon-limited NEP of $2.4\times 10^{-19}\,\,{\rm W/\sqrt{Hz}}$
calculated with noise modifier $\gamma_\phi=0.5$.
 
Further measurements are in progress to assess uniformity
of the characteristics across the array, 
to measure in detail the dark NEP and response times as a function of absorber
geometry, and to use the frame heaters 
to assess wafer heat sinking and
pixel-to-pixel thermal cross-talk. We are also 
developing a fully micro-machined wafer backing-plate incorporating 
metallized back shorts and wiring fanout for integration into a full focal-plane.
Results will be reported separately. The results already obtained
suggest an important step-forward in the development of the next generation of space-based
TESs and from the perspective of fabrication a significant step. 
\section{Acknowledgments }
This work was supported in part by  ESA Technology Research Programme
Contract No. 22359/09/NL/CP. We are  also very grateful  to colleagues working on that contract 
within the Astronomy Instrumentation Group, Cardiff
University, the  
Space Research Organization Netherlands, the National University of Ireland, Maynooth 
and the Space Science and Technology Department of Rutherford Appleton Laboratory for numerous stimulating discussions.


\end{document}